\documentclass{siamltex} % for ming
\usepackage{citesort}
\usepackage{epsfig}

\title{Provably Fast and Accurate Recovery of Evolutionary Trees 
through
Harmonic Greedy Triplets\thanks{Department of Computer Science, Yale
University, New Haven, CT 06520; $\{$csuros-miklos,
kao-ming-yang$\}$@cs.yale.edu.  Research supported in part by NSF
Grant 9531028.}}

\author{Mikl\'os Cs\H{u}r\"os \and\ Ming-Yang Kao}

\newcommand{\eqdef}{\stackrel{\rm def}{=}}
\newcommand{\newloglike}[2]{\newcommand{#1}{\mathop{\rm #2}\nolimits}}
\newloglike{\polylog}{polylog}
\newcommand{\etal}[0]{{\it et al.}}

\newcommand{\probgen}[4]{#1\!\left#2\,{#4}\,\right#3}
\newcommand{\conditional}[6]{%
    #1\!\left#2\,{#5}\,\vphantom{#6}\right#3\left.%
    \vphantom{#5}{#6}\,\right#4}
\newcommand{\bbrd}[1]{\mbox{\rm{I}\kern-.1667em{#1}}}
\newcommand{\EXP}{\bbrd{E}}
\newcommand{\PROB}{{\rm Pr}}
\newcommand{\Prob}[1]{\probgen{\PROB}{\{}{\}}{#1}}
\newcommand{\Probc}[2]{\conditional{\PROB}{\{}{|}{\}}{#1}{#2}}
\newcommand{\Exp}[1]{\probgen{\EXP}{[}{]}{#1}}
\newcommand{\tuple}[1]{\langle{#1}\rangle}

\newcommand{\AAA}{{\cal A}}
\newcommand{\BB}{{\cal B}}
\newcommand{\CC}{{\cal C}}

\newcommand{\mytab}[0]{\hspace{1\labelsep}}
\newcommand{\mt}[0]{\mytab\ }
\newcommand{\mtt}[0]{\mytab\ \mytab\ }
\newcommand{\mttt}[0]{\mytab\ \mytab\ \mytab\ }

\newlength{\heightone}
\newlength{\widthtwo}
\newlength{\widththree}
\newlength{\heightthree}
\newlength{\widfour}
\newlength{\heifour}
\newcommand{\eqref}[1]{{\textrm(\ref{#1})}}

\newcommand{\bigO}[1]{{\cal O}\left(#1\right)}
\newcommand{\smallbigO}[1]{{\cal O}(#1)}

\newcommand{\closesym}{\sigma}
\newcommand{\close}[1]{\closesym_{#1}}
\newcommand{\hatclose}[1]{\hat\close{#1}}
\newcommand{\distsym}{\Delta}
\newcommand{\dist}[1]{\distsym_{#1}}
\newcommand{\hatdist}[1]{\hat\dist{#1}}
\newcommand{\alphabet}{{\cal A}}
\newcommand{\sym}{a}
\newcommand{\abet}{m}
\newcommand{\afact}{\alpha}
\newcommand{\closesmall}{\left(1-\afact g\right)^{2d+2}}

\newcommand{\tree}{T}
\newcommand{\toposym}{\Psi}
\newcommand{\topoof}[1]{\toposym(#1)}
\newcommand{\topo}{\topoof{\tree}}
\newcommand{\wtopo}{\toposym_{\rm w}(\tree)} 
\newcommand{\hypo}{\tree^*}
\newcommand{\cand}{{\cal S}}
\newcommand{\fHGT}{{\rm Fast-HGT}}
\newcommand{\bfHGT}{Fast-HGT}

\newcommand{\smplen}{\ell}
\newcommand{\midpoint}{\closesym_{\mathrm{md}}}
\newcommand{\closelg}{\closesym_{\mathrm{lg}}}
\newcommand{\closesm}{\closesym_{\mathrm{sm}}}
\newcommand{\deftrip}{{\rm def}}
\newcommand{\greedyevent}{{\cal E}_{\rm g}}
\newcommand{\centerevent}{{\cal E}_{\rm c}}

\newcommand{\accfactor}{c}
\newcommand{\numtaxa}{n}
\newcommand{\sametopo}{=}

\newcounter{hgt}

\begin{document}

\pagestyle{myheadings} 
\markboth{\sc cs\H{u}r\"os and kao}{\sc harmonic greedy triplets}

\maketitle

\begin{abstract} 
We give a greedy learning algorithm for reconstructing an evolutionary
tree based on a certain harmonic average on triplets of terminal taxa.
After the pairwise distances between terminal taxa are estimated from
sequence data, the algorithm runs in $\smallbigO{\numtaxa^2}$ time using
$\smallbigO{\numtaxa}$ work space, where $\numtaxa$ is the number of
terminal taxa.  These time and space complexities are optimal in the
sense that the size of an input distance matrix is $\numtaxa^2$ and the size
of an output tree is $\numtaxa$.  Moreover, in the Jukes-Cantor model
of evolution, the algorithm 
recovers the correct tree topology with
high probability using sample sequences of length 
polynomial in (1) $\numtaxa$, 
(2) the logarithm of
the error probability, and (3) the inverses of two  small
parameters.
\end{abstract}



\section{Introduction} 
Algorithms for reconstructing evolutionary trees are useful tools in
biology~\cite{gus97,SOWH96}.  These algorithms
usually compare aligned character sequences for the terminal taxa in
question to infer their evolutionary relationships.  In
the past, such characters were often categorical variables of
morphological features; newer studies have taken
advantage of available biomolecular sequences.  This
paper focuses on datasets of the latter type.

We present a new learning algorithm, called {\it Fast Harmonic Greedy
Triplets} (\fHGT), using a greedy strategy based on a certain harmonic
average on triplets of terminal taxa.  After the pairwise distances
between terminal taxa are estimated from their observed sequences,
\fHGT\ runs in $\smallbigO{\numtaxa^2}$ time using
$\smallbigO{\numtaxa}$ work space, where $\numtaxa$ is the number of
terminal taxa.  These time and space complexities are optimal in the
sense that $\numtaxa^2$ is the size of an input distance matrix and
$\numtaxa$ is the size of an output tree.  An earlier variant of 
\fHGT\ takes $\smallbigO{\numtaxa^5}$ time
\cite{kao.tree.dna.soda}.  In the Jukes-Cantor model of sequence
evolution  generalized for an arbitrary
alphabet \cite{SOWH96}, \fHGT\ is proven to recover the
correct topology with high probability while requiring sample
sequences of length $\smplen$ polynomial in (1) $\numtaxa$, (2) the logarithm of
the error probability, and (3) the inverses of two small
parameters (Theorem~\ref{new_tm:sample}).  
In subsequent work \cite{csuros2000}, \fHGT\ and
its variants are shown to have similar theoretical performance in more
general Markov models of evolution.

Among the related work, there are four other algorithms which have
essentially the same guarantee on the length $\smplen$ of sample
sequences. These are the Dyadic Closure Method (DCM) \cite{essw-rsa}
and the Witness-Antiwitness Method (WAM) \cite{essw-tcs} of Erd\H{o}s,
Steel, Sz{\'e}kely, and Warnow, the algorithm of Cryan, Goldberg, and
Goldberg (CGG) \cite{CrGoGo98}, and the DCM-Buneman algorithm of Huson,
Nettles, and Warnow \cite{HNW1999}.  Not all of these results analyzed
the space complexity.  In terms of time complexity, DCM-Buneman
is not a polynomial-time algorithm.  
CGG runs in polynomial time, whose
degree has not been explicitly determined but which appears to be
higher than $\numtaxa^2$.  DCM takes
$\smallbigO{\numtaxa^5\log\numtaxa}$ time to assemble
$\smallbigO{\numtaxa^4}$ quartets using $\smallbigO{\numtaxa^4}$
space.  The two versions of WAM take
$\smallbigO{\numtaxa^6\log\numtaxa}$ and
$\smallbigO{\numtaxa^4\log\numtaxa\log\smplen}$ time, respectively.
In the uniform and Yule-Harding models of randomly generating trees,
with high probability, these two latter running times are reduced to
$\smallbigO{\numtaxa^3\polylog\numtaxa}$ and
$\smallbigO{\numtaxa^2\polylog\numtaxa}$, respectively.  Under these
two tree distributions, Erd\H{o}s \etal\ \cite{essw-rsa} further showed that
with high probability, the required sample size of DCM 
is polylogarithmic in $\numtaxa$; this bound also
applies to WAM, CGG, DCM-Buneman, and \fHGT.

Among the algorithms with no known comparbale guarantees on $\smplen$,
the Neighbor Joining Method of Saitou and Nei \cite{SOWH96} runs in
$\smallbigO{\numtaxa^3}$ time and reconstructs many trees highly accurately
in practice, although the best known upper bound on its required
sample size is exponential in $\numtaxa$ \cite{Att98}.  Maximum
likelihood methods \cite{fel83,felsenstein83} are not known to achieve
the optimal required sample size as such methods are usually expected
to \cite{Siddall1998}; moreover, all their known implementations take
exponential time to find local optima, and none can find provably
global optima.  Parsimony methods aim to compute a tree that minimizes
the number of mutations leading to the observed sequences
\cite{felsenstein82}; in general, such optimization is NP-hard
\cite{DaJoSa86}.  Some algorithms strive to find an evolutionary tree
among all possible trees to fit the observed distances the best
according to some metric \cite{Ag+96}; such optimization is NP-hard
for~$L_1$ and~$L_2$ norms \cite{Day87} and for~$L_\infty$
\cite{Ag+96}.

A common goal of the above algorithms is to construct a tree with the
same topology as that of the true tree.  In contrast, the work on
PAC-learning the true tree in the $j$-State General Markov Model
\cite{steel94} aim to construct a tree which is close to the true tree
in terms of the leaf distribution in the sense of Kearns \etal\
\cite{KMRRSS1994} but which need not be the same as the true tree.
Farach and Kannan \cite{faka96} gave an
$\smallbigO{\numtaxa^2\smplen}$-time algorithm (FK) for the symmetric case
of the $2$-state model provided that all pairs of leaves have a
sufficiently high probability of being the same.  Ambainis, Desper,
Farach, and Kannan \cite{adfk97} gave a nearly tight lower bound on
$\smplen$ for achieving a given variational distance between the true
tree and the reconstructed tree.  As for obtaining the true tree, the
best known upper bound on $\smplen$ required by FK
is exponential in $\numtaxa$. CGG
\cite{CrGoGo98} also improves upon 
FK to PAC-learn in the general $2$-state model
without the symmetry and leaf similarity constraints.

The remainder of the paper is organized as follows.  Section
\ref{sec_prel} reviews the generalized Jukes-Cantor model of sequence
evolution and discusses distance-based probabilistic techniques.
Section~\ref{new_sec:algorithm} gives \fHGT. Section \ref{sec_further}
concludes the paper with some directions for further research.

\section{Model and techniques}\label{sec_prel}
Section \ref{ss:model} defines the model of evolution used in the
paper. Section~\ref{sec_prob} defines our problem of recovering
evolutionary trees from biological sequences.
Sections~\ref{ss:close_taxa} through \ref{ss:graph_tech} develop basic
techniques for the problem.

\subsection{A model of sequence evolution}\label{ss:model}
This paper employs the generalized
Jukes-Cantor model \cite{SOWH96} of sequence evolution defined
as follows.  Let $m \geq 2$ and $\numtaxa \geq 3$ be two integers.  Let
\(
\alphabet=\left\{\sym_1,\ldots,\sym_{\abet}\right\}
\)
be a finite alphabet.  An {\em evolutionary tree}~$\tree$
for~$\alphabet$ is a rooted binary tree of~$\numtaxa$ leaves with an {\em
edge mutation probability}~$p_e$ for each tree edge~$e$.  The edge
mutation probabilities are bounded away from~$0$
and~$1-\frac{1}{\abet}$, i.e., there exist~$f$ and~$g$ such that for
every edge~$e$ of~$\tree$,
\(
 0 < f \le p_e \le g <1-\frac1\abet.
\)
Given a sequence~$s_1\cdots s_{\smplen}\in\alphabet^{\smplen}$
associated with the root of~$\tree$, a set of~$\numtaxa$ {\em mutated
sequences} in $\alphabet^{\smplen}$ is generated by~$\smplen$ random
labelings of the tree at the nodes.  These $\ell$ labelings are
mutually independent.  The labelings at the $j$-th leaf give
the~$j$-th {\it mutated sequence} $s_1^{(j)}\cdots s_\ell^{(j)}$,
where the $i$-th labeling of the tree gives the $i$-th
symbols~$s_i^{(1)},\ldots, s_i^{(n)}$.  The~$i$-th labeling is carried
out from the root towards the leaves along the edges. The root is
labeled by~$s_i$. On edge~$e$, the child's label is the same as the
parent's with probability~$1-p_e$ or is different with
probability~$\frac{p_e}{\abet-1}$ for each different symbol. Such {\it
mutations} of symbols along the edges are mutually independent.

\subsection{Problem formulation}\label{sec_prob}
The {\it topology} $\topo$ of $\tree$ is the unrooted tree obtained
from $\tree$ by omitting the edge mutation probability and by
replacing the two edges $e_1$ and $e_2$ between the root and its
children with a single edge $e_0$.  Note that the leaves of $\topo$
are labeled with the same sequences as in $\tree$, but $\topo$ need
not be labeled otherwise.  The {\it weighted topology} $\wtopo$ of
$\tree$ is $\topo$ where each edge $e \neq e_0$ of $\topo$ is further
weighted by its edge mutation probability $p_e$ in $\tree$ and for
technical reasons, the edge $e_0$ is weighted by
$1-(1-p_{e_1})(1-p_{e_2})$.  

For technical convenience, the weight of each edge $XY$ in $\wtopo$ is
often replaced by a certain edge length, such as $\dist{XY}$ in
Equation~\eqref{eq:dist}, from which the weight of $XY$ can be
efficiently determined.

The {\it weighted evolutionary topology} problem is that of taking~$\numtaxa$
mutated sequences as input and recovering $\wtopo$ with high accuracy
and high probability. \fHGT\ is a learning algorithm for
this problem.

{\it Remark.} The special treatment for $e_1$ and $e_2$ is due to the
fact that the root sequence may be entirely arbitrary and thus, in
general, no algorithm can place the root accurately.  This is
consistent with the fact that the root sequence is not directly
observable in practice, and locating the root requires considerations
beyond those of general modeling~\cite{SOWH96}. If the root sequence
is also given as input, \fHGT\ can be modified to locate the root and
the weights of $e_1$ and $e_2$ in a straightforward manner.

\subsection{Probabilistic closeness}\label{ss:close_taxa}
\fHGT\ is based on a notion of probabilistic closeness
between nodes.  For the $i$-th random labeling of $\tree$, we identify
each node of~$\tree$ with the random variable $X_i$ that gives the
labeling at the node. Note that since $s_1 \cdots s_{\smplen}$ may be
arbitrary, the random variables $X_i$ for different $i$ are not
necessarily identically distributed. For brevity, we often omit the
index $i$ of $X_i$ in a statement if the statement is independent of
$i$.

For nodes~$X$ and~$Y \in \tree$, let
\(
p_{XY} =  \Prob{X \ne Y}.
\)
The {\em closeness} of $X$ and $Y$ is
\begin{equation}\label{def:close}
\close{XY} = \Prob{X=Y}-\frac{1}{\abet-1}\Prob{X\ne Y} 
= 1 - \alpha p_{XY}, \mbox{where}\ 
\afact =  \frac{\abet}{\abet-1}.
\end{equation}

\begin{lemma}[folklore]\label{lm:exp_close}
If node~$Y$ is on the path between two nodes $X$ and~$Z$ in $\tree$,
then $ \close{XZ} = \close{XY}\close{YZ}$.
\end{lemma}

If $X$ and $Y$ are leaves, their closeness is estimated from 
sample sequences as
\begin{equation}\label{eq:est_closeness}
\hatclose{XY} = \frac{1}{\smplen}\sum_{i=1}^{\smplen}
I_{\hat{X}_i\hat{Y}_i},
\end{equation}
where~$\hat{X}_1,\ldots,\hat{X}_{\smplen}$
and~$\hat{Y}_1,\ldots,\hat{Y}_{\smplen}$ are the symbols at
positions~$1,\ldots,\smplen$ of the observed sample sequences for the
two leaves, and
\[
I_{xy} = \left\{\begin{array}{ll}
	\frac{-1}{\abet-1} & \mbox{ if } x\ne y;\\
	1 & \mbox{ if } x=y.\end{array}\right.
\]
The next lemma is useful for analyzing the estimation given by 
Equation~\eqref{eq:est_closeness}.
\begin{lemma}\label{lm:che_distance}
For~$\epsilon>0$,
\begin{eqnarray}
\label{eq:close_fraction_less}
\Prob{\frac{\hatclose{XY}}{\close{XY}} \le  1-\epsilon} 
& \le & \exp\left(-\frac{2}{\afact^2}\smplen\close{XY}^2\epsilon^2\right);
\\
\label{eq:close_fraction_more}
\Prob{\frac{\hatclose{XY}}{\close{XY}} \ge 1+\epsilon} 
& \le & \exp\left(-\frac{2}{\afact^2}\smplen\close{XY}^2\epsilon^2\right).
\end{eqnarray}
\end{lemma}
\begin{proof}
By Equation~\eqref{eq:est_closeness},
\begin{eqnarray*}
\Prob{\frac{\hatclose{XY}}{\close{XY}} \le 1-\epsilon} 
& = & \Prob{\sum_{i=1}^{\smplen} 
\left(I_{X_iY_i}-\close{XY}\right) \le -\smplen\close{XY}\epsilon};
\\
\Prob{\frac{\hatclose{XY}}{\close{XY}} \ge 1+\epsilon} 
& = & \Prob{\sum_{i=1}^{\smplen} 
\left(I_{X_iY_i}-\close{XY}\right) \ge \smplen\close{XY}\epsilon}.
\end{eqnarray*} 
Since~$\frac{-1}{m-1}\le I_{X_iY_i}\le 1$
and~$\Exp{I_{X_iY_i}-\close{XY}}=0$, we use Hoeffding's
inequality~\cite{Hoe63}~on sums of independent bounded random
variables to have Equations~\eqref{eq:close_fraction_less}
and~\eqref{eq:close_fraction_more}.
\end{proof}

\subsection{Distance and harmonic mean}\label{sec_mean}
The {\it distance} of nodes~$X$ and~$Y \in \tree$~is
\begin{equation}\label{eq:dist}
\dist{XY} = -\ln\close{XY}.
\end{equation}
For an edge $XY$ in $\tree$, $\dist{XY}$ is called the {\em edge
length} of $XY$.

\fHGT\ uses Statement \ref{lem_somepoint} of the next corollary 
to locate internal nodes of $\tree$.
\begin{corollary}\label{cor_dist}
Let $X$, $Y$, and $Z$ be nodes in $\tree$.
\begin{enumerate}
\item \label{cor_dist_1}
If $X \neq Y$, then $\dist{XY}=\dist{YX} > 0$. Also,
$\dist{XX}=0$.
\item \label{cor_dist_add} If $Y$ is on the path between~$X$ and~$Z$
in $\tree$, then $\dist{XZ}=\dist{XY}+\dist{YZ}$.
\item\label{lem_somepoint} 
For any $\closesym$ with
$\close{XY}\le\closesym < 1$, there is a node~$P$ on the path between~$X$
and~$Y$ in $\tree$ such that \(\closesym(1-\afact g)^{1/2}
\le\close{XP} \le\closesym(1-\afact g)^{-1/2}.\)
Furthermore, if \(\close{XY}(1-\afact g)^{1/2}
<\closesym < (1-\afact g)^{-1/2},\) then
$P$ is distinct from $X$ and $Y$.
\end{enumerate}
\end{corollary}
\begin{proof} Statements~\ref{cor_dist_1} and \ref{cor_dist_add}
follow from Equation~\eqref{def:close} and
Lemma~\ref{lm:exp_close}.
Statement~\ref{lem_somepoint} 
becomes straightfoward when restated in 
terms of distance as follows.
For any $\distsym$ with
$\dist{XY}\ge\distsym>0$, there is a node~$P$ on the path between~$X$
and~$Y$ in $\tree$ such that \(\distsym+\frac{-\ln(1-\afact g)}{2}
\ge\dist{XP} \ge\distsym-\frac{-\ln(1-\afact g)}{2}.\)
Furthermore, if \(\dist{XY}-\frac{-\ln(1-\afact g)}{2}
> \distsym > \frac{-\ln(1-\afact g)}{2},\) then
$P$ is distinct from $X$ and $Y$.
\end{proof}

If $X$ and $Y$ are leaves, their distance is estimated from sample
sequences as
\begin{equation}\label{eq:est_distance}
\hatdist{XY} = \left\{\begin{array}{ll}
	-\ln\hatclose{XY} & \mbox{ if }\hatclose{XY}>0;\\
	\infty & \mbox{ otherwise.}\end{array}\right.
\end{equation}

\begin{figure}[h]
\centerline{\psfig{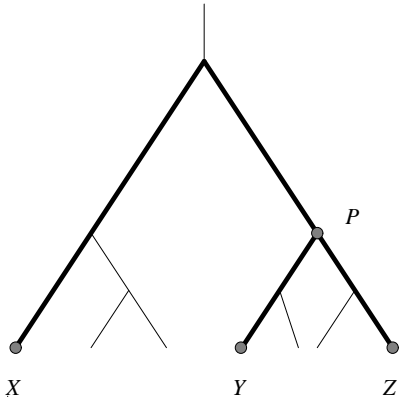}}
\caption{$P$ is the center of triplet $XYZ$.}
\label{fig:center_triplet}
\end{figure}

A {\em triplet} $XYZ$ consists of three distinct leaves $X$, $Y$, and
$Z$ of $\tree$.  There is an internal node $P$ in $\tree$ at which the
pairwise paths between the leaves in $XYZ$ intersect; see
Figure~\ref{fig:center_triplet}. $P$ is the {\em center} of $XYZ$, and
$XYZ$ {\em defines} $P$. Note that a star is formed by the edges on
the paths between $P$ and the three leaves in $XYZ$.

By Corollary~\ref{cor_dist}(\ref{cor_dist_add}),
the distance between $P$ and a leaf in $XYZ$, say,
$X$, can be obtained as
\(
\dist{XP}=\frac{\dist{XY}+\dist{XZ}-\dist{YZ}}{2},
\)
which is estimated by
\begin{equation}\label{eq:est_center}
\hatdist{XP}=\frac{\hatdist{XY}+\hatdist{XZ}-\hatdist{YZ}}{2}.
\end{equation}

The {\em closeness} of $XYZ$ is 
\(
\close{XYZ} = \frac{3}{\frac {1}{\close{XY}} +\frac{1}{\close{XZ}}
	+\frac{1}{\close{YZ}}},
\)
which is estimated by
\(
\hatclose{XYZ} = \frac{3}{\frac {1}{\hatclose{XY}} +
	\frac{1}{\hatclose{XZ}} +\frac{1}{\hatclose{YZ}}}.
\)
$XYZ$ is called {\it positive} if $\hatclose{XY}$, $\hatclose{XZ}$,
and $\hatclose{YZ}$ are all positive.

The next corollary relates $\close{XYZ}$ and the pairwise closenesses
of $X$, $Y$, and $Z$.
\begin{corollary}\label{eq:avg} 
If $\close{XP}\le\close{YP}\le\close{ZP}$, then
$\close{XY}\le\close{XZ}\le\close{YZ}$,
$\close{XZ}\ge\frac{2}{3}\close{XYZ}$, and $\close{YP}^2 \ge
\frac{1}{3}\close{XYZ}$.
\end{corollary}
\begin{proof}
This corollary follows from Lemma~\ref{lm:exp_close}
and simple algebra.
\end{proof}

The next lemma relates $\close{XYZ}$ to the probability of
overestimating the distance between $P$ and a leaf in $XYZ$ using
Equation~\eqref{eq:est_center}.
\begin{lemma}\label{lm:est_center} For~$0<\epsilon<1$, 
\[
\Prob{\hatdist{XP}-\dist{XP} \ge \frac{-\ln(1-\epsilon)}{2}}
\le 3\exp\left(-\frac{2}{9\afact^2}\smplen\close{XYZ}^2\epsilon^2\right).
\]
\end{lemma}
\begin{proof}
See \S\ref{app:lm:est_center}.
\end{proof}

\subsection{Basis of a greedy strategy}\label{ss:graph_tech} Let
$d_{XY}$ denote the number of edges in the  path between two
leaves $X$ and~$Y$ in $\tree$.  By Lemma~\ref{lm:exp_close},
$\close{XY}$ can be as small as $(1-\afact g)^{d_{XY}}$.  Thus, the
larger $d_{XY}$ is, the more difficult it is to estimate $\close{XY}$
and $\dist{XY}$. This intuition leads to a natural greedy strategy
outlined below that favors leaf pairs with small $d_{XY}$ and large
$\close{XY}$.

The {\it g-depth} of a node in a rooted tree  $T'$ is the smallest number of
edges in a path from the node to a leaf.  Let $e$ be an edge between
nodes $u_1$ and $u_2$.  Let $T'_1$ and $T'_2$ be
the subtrees of $T'$ obtained by cutting $e$ which contain $u_1$ and
$u_2$, respectively.  The {\em g-depth} of $e$ in $T'$ is the larger
of the g-depth of $u_1$ in $T'_1$ and that of $u_2$ in $T'_2$.  The
{\em g-depth} of a rooted tree is the largest possible g-depth of an
edge in the tree. (The prefix g emphasizes that this usage of depth is
nonstandard in graph theory.)

Let $d$ be the $g$-depth of $\tree$.  Variants of the next lemma have
proven very useful and insightful; see, e.g.,
\cite{DuZhFe91,essw-tcs,essw-rsa}.
\begin{lemma}\
\label{fact:g-things}

\begin{enumerate}
\item \(d \leq 1+\lfloor{\log_2 (n-1)}\rfloor.\)
\item\label{fact:g:large} Every internal node $P$ of $\tree$ except
the root has a defining triplet $XYZ$ such that $d_{XP}, d_{YP}$, and
$d_{ZP}$ are all at most $d+1$ and thus, $\close{XYZ} \geq (1-\afact
g)^{2(d+1)}$. Every leaf of $\tree$ is in such a triplet.
\end{enumerate}
\end{lemma}
\begin{proof}
The proof is straightforward.  Note that the more unbalanced $\tree$
is, the smaller its g-depth is.  
\end{proof}

In $\tree$, the star formed by a defining triplet of an internal node
contains the three edges incident to the internal node.  Thus, $\topo$
can be reconstructed from triplets described in
Lemma~\ref{fact:g-things}(\ref{fact:g:large}) or those with similarly
large closenesses.  This observation motivates the following
definitions.  Let
\[
\closelg  = \frac{3\sqrt{2}}{2}\left(\frac{\sqrt{2}-1}{\sqrt{2}+1}\right)^2
(1-\afact g)^{2d+4};\ \ 
\closesm  =  \frac{\closelg}{\sqrt{2}};\ \ 
\midpoint =  \frac{\closelg+\closesm}{2}.
\]
{\it Remark.}  The choice of $\closelg$ is obtained by solving
Equations~\eqref{new_eq:close_large_one}, \eqref{new_eq:close_large_two}
and~\eqref{new_eq:close_large_three}.

A triplet $XYZ$ is {\it large} if $\close{XYZ} \geq \closelg$; it is
{\em small} if $\close{XYZ} \leq \closesm$.  Note that by
Lemma~\ref{fact:g-things}(\ref{fact:g:large}), each nonroot internal
node of $\tree$ has at least one large defining triplet.  

\begin{lemma}\label{lm:greedy}
The first inequality below holds for all large triplets $XYZ$, and the
second for all small triplets.
\begin{eqnarray}\label{eq:greedy_large}
\Prob{\hatclose{XYZ}  \le \midpoint} 
& \le & \exp\left( -\frac{\left(\sqrt{2}-1\right)^2}{36\alpha^2}
\smplen\closelg^2\right);  
\\
\label{eq:greedy_small}
\Prob{\hatclose{XYZ} \ge \midpoint}
& \le & \exp\left( -\frac{\left(\sqrt{2}-1\right)^2}{36\alpha^2}
\smplen\closelg^2\right).
\end{eqnarray}
\end{lemma}
\begin{proof}
See \S\ref{app:lm:greedy}.
\end{proof}

A nonroot internal node of $\tree$ may have more than one large
defining triplet. Consequently, since distance estimates contain
errors, we may obtain an erroneous estimate of~$\topo$ by
reconstructing the same internal node more than once from its
different large defining triplets. To address this issue, \fHGT\
adopts a threshold
\(
0 < \dist{\min} < \frac{-\ln(1-\afact f)}{2}
\)
based on the fact that the distance between two distinct nodes is at
least $-\ln(1-\afact f)$; also let
\(
\accfactor = \frac{\dist{\min}}{-\ln(1-\afact f)}.
\)
\fHGT\ considers the center~$P$ of a triplet~$XYZ$ and
the center~$Q$ of another triplet~$XUV$ to be separate if and only if
\begin{eqnarray}\label{endpoint_compare}
|\hatdist{XP}-\hatdist{XQ}|\ge\dist{\min},\  
\end{eqnarray}
where
\(
\hatdist{XP}=(\hatdist{XY}+\hatdist{XZ}-\hatdist{YZ})/2\ \mbox{and}\ 
\hatdist{XQ} = (\hatdist{XU}+\hatdist{XV}-\hatdist{XV})/2.
\)
Notice that two triplet centers can be compared in this manner 
only if the triplets share at least one leaf. 
The next lemma shows that a large triplet's
center is estimated within a small error with high probability.
\begin{lemma}\label{lm:center}
Let $P$ be the center of a triplet~$XYZ$.  If $XYZ$ is not small, then
\begin{eqnarray}\label{eq:center_ok}
\Prob{\left|\hatdist{XP}-\dist{XP}\right| 
				\ge \frac{\dist{\min}}{2}}
& \le & 7\exp\left(-\frac{\accfactor^2}{81}\smplen\closelg^2 f^2\right).
\end{eqnarray}
\end{lemma}
\begin{proof}
See \S\ref{app:lm:center}.
\end{proof}

We next define and analyze two key events $\centerevent$ and
$\greedyevent$ as follows.  The subscripts c and g denote the words
greedy and center, respectively.
\begin{itemize}
\item 
$\centerevent$ is the event that for every triplet~$XYZ$ that
is not small, $\left|\hatdist{XP}-\dist{XP}\right| <
\frac{\dist{\min}}{2}$, $\left|\hatdist{YP}-\dist{YP}\right| <
\frac{\dist{\min}}{2}$, and $\left|\hatdist{ZP}-\dist{ZP}\right| <
\frac{\dist{\min}}{2}$, where $P$ is the center of $XYZ$.  
\item 
$\greedyevent$ is the event that $\hatclose{XYZ}>\hatclose{X'Y'Z'}$
for every large triplet~$XYZ$ and every small triplet $X'Y'Z'$. 
\end{itemize}

\begin{lemma}\label{lem_events}\

\[
\Prob{\overline\centerevent}  \le  21{n\choose 3}
	\exp\left(-\frac{\accfactor^2}{81}\smplen\closelg^2 f^2\right); \
\Prob{\overline\greedyevent}  \le 
{n\choose 3} \exp\left( -\frac{\left(\sqrt{2}-1\right)^2}{36\alpha^2}
\smplen\closelg^2\right).
\]
\end{lemma}
\begin{proof} 
The inequalities follow from
Equation~\eqref{eq:center_ok}~and~Lemma~\ref{lm:greedy}, respectively.
\end{proof}

\section{\bfHGT}\label{new_sec:algorithm}
Section~\ref{new_ss:build_tree} details \fHGT.
Section~\ref{new_sec_time} analyzes its running time and work space.
Section \ref{new_ss:run_sample} proves technical lemmas for bounding
the algorithm's required sample size.  Section
\ref{new_ss:main_sample} analyzes this sample size.

\subsection{The description of \bfHGT}\label{new_ss:build_tree}

\begin{figure}[h]
\begin{list}{\thehgt}
{\usecounter{hgt}\setcounter{hgt}{0}\renewcommand{\thehgt}{F\arabic{hgt}}
\setlength{\rightmargin}{0in} 
\settowidth{\leftmargin}{F29}\addtolength{\leftmargin}{\labelsep}}
\item[]
{\bf Algorithm} Fast Harmonic Greedy Triplets
\item[] {\bf Input:} 
\begin{itemize}
\item $\dist{\min}$;
\item 
$\hatdist{XY}$ for all leaves $X$ and $Y$ of $\tree$ which are
computed via Equations \eqref{eq:est_closeness} and
\eqref{eq:est_distance} from $\numtaxa$ mutated length-$\smplen$
sequences generated by $\tree$.
\end{itemize}
\item[]
{\bf Output:} $\wtopo$.
\item\label{new_fhgt:init_triplet}
Select an arbitrary leaf $A$ and find a triplet $ABC$ with the maximum
$\hatclose{ABC}$.
\item\label{new_hgt_fail_1}
{\bf if} $AB$C is not positive {\bf then} let $\hypo$ be the empty
tree, {\bf fail}, and {\bf stop}.
\item\label{new_line:init_tree} 
Let $\hypo$ be the star with three edges formed by $ABC$ and its
center $D$.
\item\label{new_fhgt:init_length}
Use Equation \eqref{eq:est_center} to set
$\dist{AD}^*\leftarrow\hatdist{AD}$,
$\dist{BD}^*\leftarrow\hatdist{BD}$,
$\dist{CD}^*\leftarrow\hatdist{CD}$.
\item\label{new_line:star} Set $\deftrip(D)\leftarrow\{A, B, C\}$.
\item\label{new_find_update_0}
First set all $\cand[M]$ to null; then for $Q_1Q_2
\in \{AD,BD,CD\}$, Update-$\cand$$(Q_1Q_2)$.
\item\label{new_fhgt:repeat}
{\bf repeat}
\item \mt\  \label{new_hgt_fail_2}
{\bf if} $\cand[M]=$ null for all leaves $M \in \tree$ {\bf then} {\bf fail}
and {\bf stop}.
\item \mt\ \label{new_fhgt:choose}
Find
$\cand[N]=\tuple{P_1P_2,NXY,P,\dist{P_1P}^*,\dist{P_2P}^*,\dist{NP}^*}$
with the maximum $\hatclose{NXY}$.
\item \mt\  \label{new_line:add_int} 
Split $P_1P_2$ into two edges $P_1P$ and $P_2P$ in $\hypo$ with
lengths $\dist{P_1P}^*$ and $\dist{P_2P}^*$.
\item \mt\  \label{new_line:add_leaf}
Add to $\hypo$ a leaf $N$ and an edge $NP$ with length $\dist{NP}^*$.
\item \mt\  \label{new_line:add_split}
Set $\deftrip(P)\leftarrow\{N,X,Y\}$.
\item \mt\  \label{new_find_update_2}\label{new_line:delete}
For every $M$ with~$\cand[M]$ containing the edge $P_1P_2$, set
$\cand[M]\leftarrow$ null.
\item \mt\ \label{new_find_update_3}\label{new_line:add_new}
For each $Q_1Q_2 \in \{P_1P,P_2P,NP\}$, Update-$\cand$$(Q_1Q_2)$.
\item 
{\bf until} all leaves of $\tree$ are inserted to $\hypo$; i.e., this
loop has iterated $\numtaxa-3$ times.
\item \label{new_line_output} Output $\hypo$.
\end{list}
\caption{The \fHGT\ algorithm.}
\label{new_fig_fhgt}
\end{figure}

\begin{figure}[h]
\begin{list}{\thehgt}
{\usecounter{hgt}\setcounter{hgt}{0}\renewcommand{\thehgt}{U\arabic{hgt}}
\setlength{\rightmargin}{0in} 
\settowidth{\leftmargin}{A9}\addtolength{\leftmargin}{\labelsep}}
\item[] 
{\bf Algorithm} Update-$\cand$ 
\item[] {\bf Input:} an edge  $Q_1Q_2 \in \hypo$
\item\label{new_update:find} 
Find all splitting tuples for $Q_1Q_2 \in \hypo$.
\item\label{new_update:best}
For each
$\tuple{Q_1Q_2,MUV,Q,\dist{Q_1Q}^*,\dist{Q_2Q}^*,\dist{MQ}^*}$ at line
\ref{new_update:find}, assign it to $\cand[M]$ if $\hatclose{MUV}$ is greater than
that of $\cand[M]$.
\end{list}
\caption{The Update-$\cand$ subroutine.}
\label{new_fig_updateS}
\end{figure}

\begin{figure}[h]
\begin{list}{\thehgt}
{\usecounter{hgt}\setcounter{hgt}{0}\renewcommand{\thehgt}{S\arabic{hgt}}
\setlength{\rightmargin}{0in}
\settowidth{\leftmargin}{S27}\addtolength{\leftmargin}{\labelsep}}
\item[]
{\bf Algorithm} Split-Edge
\item[]
{\bf Input:} an edge $P_1P_2$ in $\hypo$ and a relevant triplet $NXY$
with center $P$.
\item[]
{\bf Output:} If $P$ is strictly between $P_1$ and $P_2$ in $\tree$
and thus can be inserted on $P_1P_2$, then we return the message
``split'' and the edge lengths $\dist{P_1P}^*$, $\dist{P_2P}^*$, and
$\dist{NP}^*$.  Otherwise, we return a reason why $P$ cannot be
inserted.
\item
Use Equation \eqref{eq:est_center} to compute $\hatdist{XP}$,
$\hatdist{YP}$, $\hatdist{NP}$ for  $NXY$.
\item 
Let $X_1\in\{X,Y\}\cap\deftrip(P_1)$ and
$X_2\in\{X,Y\}\cap\deftrip(P_2)$.
\item\label{new_line:delta_xp_one}
For each $i = 1$ or $2$, 
{\bf if} $P_i$ is an internal node of $\hypo$
\item \mt\ 
{\bf then} use Equation \eqref{eq:est_center} to compute
$\hatdist{X_iP_i}$ for the triplet formed by $\deftrip(P_i)$
\item \mt\ {\bf else} set $\hatdist{X_iP_i}\leftarrow 0$.
\item\label{new_line:delta_one_two}
Set $\distsym_1 \leftarrow\hatdist{X_1P}-\hatdist{X_1P_1}$ and
$\distsym_2  \leftarrow\hatdist{X_2P}-\hatdist{X_2P_2}$.
\item \label{new_line:separate_test}
{\bf if} $|\distsym_1|<\dist{\min}$ or $|\distsym_2|<\dist{\min}$
\item \mt\ \label{new_line:separate_test_too_close}
{\bf then} {\bf return} ``too close''
\item \mt\ {\bf else} {\bf begin}
\item \mtt\ \label{new_line:between_1}
{\bf if} $P_2$ (respectively, $P_1$) is on the path between $P_1$ and $X_1$ 
($P_2$ and $X_2$)
in $\hypo$
\item \mttt\ {\bf then} set $\distsym_1'\leftarrow -\distsym_1$ 
($\distsym_2' \leftarrow -\distsym_2$) 
\item \mttt\ {\bf else} set $\distsym_1'\leftarrow \distsym_1$ ($\distsym_2' \leftarrow \distsym_2$).
\item[] \mtt\ ({\it Remark.} Since $X_1$ may equal $X_2$, 
the tests for $P_1$ and $P_2$ are both needed.)
\item \mtt\  \label{new_line:set_length_one}
Set $\distsym_1'' \leftarrow (\distsym_1'+\dist{P_1P_2}^*-\distsym_2')/2$ and
$\distsym_2'' \leftarrow (\distsym_2'+\dist{P_1P_2}^*-\distsym_1')/2$.

\mtt\ ({\it Remark.} $\distsym_1''+\distsym_2'' = 
\dist{P_1P_2}^*$, $\distsym_1''$ estimates $\dist{P_1P}$, and 
$\distsym_2''$ estimates $\dist{P_2P}$.)

\item \mtt\ \label{new_line:test_length} 
{\bf if} $\distsym_1'' \geq \dist{P_1P_2}^*$ or $\distsym_2'' \geq
\dist{P_1P_2}^*$ 
\item \mttt\ \label{new_split_outside}
{\bf then} {\bf return} ``outside this edge''
\item \mttt\ \label{new_split_add}
{\bf else} {\bf return} ``split'', $\distsym_1''$, $\distsym_2''$,
$\hatdist{NP}$.
\item \mt\ {\bf end.}
\end{list}
\caption{The Split-Edge subroutine.}
\label{new_fig_split}
\end{figure}

\fHGT\ and its subroutines Update-$\cand$  and Split-Edge are detailed in
Figures~\ref{new_fig_fhgt}, \ref{new_fig_updateS}, and
\ref{new_fig_split}, respectively.

Given $\dist{\min}$ and $\numtaxa$ mutated sequences as input, the
task of \fHGT\ is to recover $\wtopo$.  The algorithm first constructs
a star $\hypo$ formed by a large triplet at
lines~\ref{new_fhgt:init_triplet} through \ref{new_line:init_tree}.
It then inserts into $\hypo$ a leaf of $\tree$ and a corresponding
internal node per iteration of the repeat at
line~\ref{new_fhgt:repeat} until ~$\hypo$ has a leaf for each input
sequence. The $\hypo$ at line \ref{new_line_output} is our
reconstruction of $\wtopo$.  For $k = 3,\ldots,\numtaxa$,
let~$\hypo_k$ be the version of $\hypo$ with~$k$ leaves constructed
during a run of
\fHGT; i.e., $\hypo_3$ is constructed at line
\ref{new_line:init_tree}, and $\hypo_k$ with $ k \geq 4$ is constructed at
line \ref{new_line:add_leaf} during the $(k-3)$-th iteration of the
repeat. Note that $\hypo_n$ is output at line \ref{new_line_output}.

%\here{10/22/00 -- 18:00}

A node $Q$ is {\it strictly between} nodes $Q_1$ and $Q_2$ in $\tree$
if $Q$ is on the path between $Q_1$ and $Q_2$ in $\tree$ but $Q \neq
Q_1$, $Q \neq Q_2$, and $Q$ is not the root of $\tree$.  At each
iteration of the repeat, \fHGT\ finds an
edge~$P_1P_2$ in $\hypo$ and a triplet~$NXY$ where $X,Y\in\hypo$,
$N\not\in\hypo$, and the center $P$ of~$NXY$ is strictly between on
$P_1$ and $P_2$ in $\hypo$.  Such $P_1P_2$ and $NXY$ can be used to
insert $N$ and $P$ into $\hypo$. We record an insertion by letting
$\deftrip(P)=\{N,X,Y\}$; for notational uniformity, let
$\deftrip(X)=\{X\}$ for all leaves $X$.  

At line~\ref{new_find_update_0}, $\cand$ is an array indexed by the
leaves $M$ of $\hypo$. At the beginning of each iteration of the
repeat, $\cand[N]$ stores the most suitable $P_1P_2$ and $NXY$ for
inserting $N$ into $\hypo$.  $\cand$ is initialized at
line~\ref{new_find_update_0}; it is updated at lines~\ref{new_line:delete}
and \ref{new_line:add_new} after a new leaf and a new internal node
are inserted into $\hypo$.  The precise content of $\cand$ is
described in Lemma~\ref{new_lm:cand_ok}.

To further specify $\cand[N]$, we call $NXY$ {\em relevant} for
$P_1P_2\in\hypo_k$ if it is positive, $N\not\in\hypo_k$,
$X\in\deftrip(P_1)$, $Y\in\deftrip(P_2)$, and $P_1P_2$ is on the path
between~$X$ and~$Y$ in~$\hypo_k$.  We use Split-Edge to determine
whether the center~$P$ of a relevant $NXY$ is strictly between $P_1$
and $P_2$ in $\tree$.  We also use Split-Edge to calculate an
estimation $\dist{P'P''}^*$ of $\dist{P'P''}$ for each edge
$P'P''\in\hypo_k$, which is called the {\it length} of $P'P''$ in
$\hypo_k$.  Split-Edge has three possible outcomes:
\begin{enumerate}
\item 
At line~\ref{new_line:separate_test_too_close}, $P$ is too
close to $P_1$ or $P_2$ to be a different internal node. 
\item 
At line~\ref{new_split_outside}, $P$ is outside the path
between $P_1$ and $P_2$ in $\tree$ and thus should not be inserted into
$\hypo_k$ on $P_1P_2$.
\item At line~\ref{new_split_add}, $P$ is strictly between
$P_1$ and $P_2$ in $\tree$. Thus, $P$ can be inserted between $P_1$
and $P_2$ in $\hypo_k$, and the lengths
$\dist{P_1P}^*,\dist{P_2P}^*,\dist{NP}^*$ of the possible new edges
$P_1P$, $P_2P$, and~$NP$ are returned.
\end{enumerate}
In the case of the third outcome, $NXY$ is called a {\em splitting
triplet} for $P_1P_2$ in $\hypo_k$, and
$\tuple{P_1P_2,NXY,P,\dist{P_1P}^*,\dist{P_2P}^*,\dist{NP}^*}$ is a
{\em splitting tuple}.  
Each $\cand[N]$
is either a single splitting tuple or null. In the latter
case, the estimated closeness of the triplet in $\cand[N]$ is regarded
as $0$ for technical uniformity.

\fHGT\ ensures the accuracy of $\hypo$ in several ways.  The algorithm
uses only positive triplets to recover internal nodes of $\tree$ at
lines \ref{new_fhgt:init_triplet} and \ref{new_fhgt:choose}.  These
two lines together form the greedy strategy of \fHGT.  The maximality
of the triplet chosen at these two lines favors large triplets over
small ones based on Lemmas~\ref{fact:g-things} and \ref{lm:greedy}.
With a relevant triplet as input, Split-Edge compares $P$ to $P_1$ and
$P_2$ using the rule of Equation~\eqref{endpoint_compare} and can
estimate the distance between $P$ and $P_1$ or $P_2$ from the same
leaf to avoid accumulating estimation errors in edge lengths.

The next lemma enables \fHGT\ to grow $\hypo$ by always using relevant
triplets.  
\begin{lemma}\label{new_lem:cand_correct_endpoint}
For each $k=3,\ldots,\numtaxa-1$, at the start of the $(k-2)$-th
iteration of the repeat at line~\ref{new_fhgt:repeat},
$\deftrip(P_1) \cap
\deftrip(P_2)\ne\emptyset$ for every edge~$P_1P_2\in \hypo_k$. 
\end{lemma}
\begin{proof}
The proof is by induction on $k$.  The base case follows from the fact
that the statement holds for $\hypo_3$ at
line~\ref{new_line:init_tree}.  The induction step follows from the
use of a relevant triplet at line~\ref{new_fhgt:choose}.
\end{proof}

{\it Remark.}  A subsequence work \cite{csuros2000} shows that \fHGT\
can run with the same time, space, and sample complexities without
knowing $f$ and $\dist{\min}$; this is achieved by slightly modifying
some parts of Split-Edge.

\subsection{The running time and work space of \bfHGT}
\label{new_sec_time}

Before proving the desired time and space complexities of \fHGT\ in
Theorem~\ref{new_tm:time_fast} below, we note the following three key
techniques used by \fHGT\ to save time and space.
\begin{enumerate}
\item
At line~\ref{new_fhgt:init_triplet}, $ABC$ is selected for a fixed
arbitrary $A$. This limits  the number of triplets considered at line
\ref{new_fhgt:init_triplet} to
$\bigO{\numtaxa^2}$.  This technique is supported by the fact that
each leaf in $\tree$ is contained in a large triplet.
\item
At lines \ref{new_find_update_0} and \ref{new_find_update_3}, $\cand$
keeps only splitting tuples. This limits the number of
triplets considered for each involved edge to $\bigO{\numtaxa}$.  This
technique is feasible since by Lemma~\ref{new_lm:int_D}, $\topo$ can
be recovered using only relevant triplets.
\item
At line \ref{new_find_update_3}, $\cand$ includes no new 
splitting tuples for the edges $Q_1Q_2$ that already exist in $\hypo$
before $N$ is inserted. This technique is feasible because the
insertion of $N$ results in no new relevant triplets for 
such $Q_1Q_2$ at all.
\end{enumerate}

\begin{theorem}\label{new_tm:time_fast}
\fHGT\  runs in $\bigO{\numtaxa^2}$
time using $\bigO{\numtaxa}$ work space.
\end{theorem}
\begin{proof}
We analyze the time and space complexities separately as follows.

{\it Time complexity.}  Line~\ref{new_fhgt:init_triplet} takes
$\bigO{\numtaxa^2}$ time.  Line~\ref{new_find_update_0}
takes~$\bigO{\numtaxa}$ total time to examine $2(\numtaxa-3)$ triplets
for each $Q_1Q_2$.  As for the repeat at line~\ref{new_fhgt:repeat},
lines~\ref{new_hgt_fail_2}, \ref{new_fhgt:choose}, and
\ref{new_find_update_2} take $\bigO{\numtaxa}$ time to search through
$\cand$.  For the $(k-3)$-th iteration of the repeat where
$k=4,\ldots,\numtaxa-1$, line~\ref{new_line:add_new} takes
$\bigO{\numtaxa}$ total time to examine at most $9(\numtaxa-k-1)$
triplets for each of $P_1P, P_2P$ and $NP$.  Thus, each iteration of
the repeat takes $\bigO{\numtaxa}$ time.  Since the repeat iterates at
most $\numtaxa-3$ times, the time complexity of \fHGT\ is as stated.

{\it Space complexity}.  $\hypo$ and the sets $\deftrip(G)$ for all
nodes $G$ in $\hypo$ take $\bigO{\numtaxa}$ work space. $\cand$ takes
$\bigO{\numtaxa}$ space.  Lines
\ref{new_fhgt:init_triplet}, \ref{new_find_update_0} and
\ref{new_line:add_new} in \fHGT\ and lines
\ref{new_update:find} and \ref{new_update:best} in Update-$\cand$
can be implemented to use $\bigO{1}$ space.  The other variables
needed by \fHGT\ take $\bigO{1}$ space. Thus, the space complexity of
\fHGT\ is as stated.  
\end{proof}

\subsection{Technical lemmas for bounding the sample size}\label{new_ss:run_sample}

Let $L_k$ be the set of the leaves of~$\topo$ that are in
$\hypo_k$. Let $\toposym_k$ be the subtree of $\topo$ formed by the edges
on paths between leaves in $L_k$.  A {\it branchless} path in~$\toposym_k$
is one whose internal nodes are all of degree~$2$ in~$\toposym_k$.  We say
that~$\hypo_k$ {\em matches}~$\tree$ if~$\hypo_k$ without the edge
lengths can be obtained from $\toposym_k$ by replacing every maximal
branchless path with an edge between its two endpoints.

For $k=3,\ldots,\numtaxa$, we define the following conditions:
\begin{itemize} 
\item $\AAA_k$: $\hypo_k$ matches $\tree$.
\item $\BB_k$: For every internal node~$Q\in\hypo_k$,
the triplet formed by $\deftrip(Q)$ is not small.
\item  $\CC_k$: For every edge~$Q_1Q_2\in\hypo_k$,
$|\dist{Q_1Q_2}^*-\dist{Q_1Q_2}| < 2\dist{\min}$.
\end{itemize} 

In this section, Lemmas~\ref{new_lm:int_one}, \ref{new_lm:int_two}, and
\ref{new_lm:int_D} analyze under what conditions Split-Edge can help
correctly insert a new leaf and a new internal node to $\hypo_k$.
Later in \S\ref{new_ss:main_sample}, we use these lemmas to show by
induction in Lemma~\ref{new_lm:sample_induct} that the
events~$\greedyevent$ and~$\centerevent$, which are defined before 
Lemma \ref{lem_events}, imply that $\AAA_k$, $\BB_k$,
and $\CC_k$ hold for all $k$. This leads to Theorem~\ref{new_tm:sample},
stating that  \fHGT\  solves the weighted evolutionary
topology problem with a polynomial-sized sample.

Lemmas~\ref{new_lm:int_one}, \ref{new_lm:int_two}, and \ref{new_lm:int_D} make the
following assumptions for some $k < \numtaxa$:
\begin{itemize}
\item
The $(k-3)$-th iteration of the repeat at line~\ref{new_fhgt:repeat}
has been completed.
\item 
$\hypo_k$ has been constructed, and $\AAA_k$, $\BB_k$, and $\CC_k$
hold.
\item 
$\fHGT$  is currently in the $(k-2)$-th iteration of the
repeat.
\end{itemize}

\begin{lemma}\label{new_lm:int_one}
Assume that $\centerevent$ holds and the triplet~$NXY$ input to
Split-Edge is not small. Then, the test of
line~\ref{new_line:separate_test} fails if and only if $P \neq P_1$ and $P
\neq P_2$ in~$\tree$.
\end{lemma}
\begin{proof}
There are two directions, both using the following equation.  By 
line \ref{new_line:delta_one_two},
\begin{eqnarray}\label{new_eq_int_one}
\distsym_1 = 
	(\hatdist{X_1P}-\dist{X_1P})-(\hatdist{X_1P_1}-\dist{X_1P_1})
	+(\dist{X_1P}-\dist{X_1P_1}).
\end{eqnarray}

$(\Longrightarrow)$ To prove by contradiction, assume $P=P_1$ or
$P=P_2$ in~$\tree$.  If $P=P_1$, then $\dist{X_1P}=\dist{X_1P_1}$, and
by $\AAA_k$, $P_1$ is an internal node in $\hypo_k$.  By $\BB_k$, the
triplet formed by $\deftrip(P_1)$ is not small. Thus, by
$\centerevent$ and Equation~\eqref{new_eq_int_one},
$|\distsym_1|<\dist{\min}$.  By symmetry, if $P=P_2$, then
$|\distsym_2|<\dist{\min}$.  In either case, the test of
line~\ref{new_line:separate_test} passes.

$(\Longleftarrow)$ 
Since $P \neq P_1$, $\dist{X_1P}-\dist{X_1P_1}\ge-\ln(1-\afact
f)\ge2\dist{\min}$.  If $P_1$ is a leaf in $\hypo_k$, then by
$\AAA_k$, $P_1$ is leaf $X_1$ in $\tree$, and
$\hatdist{X_1P_1}=\dist{X_1P_1}=0$.  By $\centerevent$ and
Equation~\eqref{new_eq_int_one}, $|\distsym_1|>1.5\dist{\min}$.  If $P_1$
is an internal node in $\hypo_k$, then by $\BB_k$, $\centerevent$, and
Equation~\eqref{new_eq_int_one}, we have $|\distsym_1|>\dist{\min}$. In
either case, $|\distsym_1|>\dist{\min}$.  By symmetry, since $P \neq
P_2$, $|\distsym_2|>\dist{\min}$.  Thus, the test of
line~\ref{new_line:separate_test} fails.
\end{proof}

\begin{lemma}\label{new_lm:int_two}
In addition to the assumption in Lemma \ref{new_lm:int_one}, also assume
that $P \neq P_1$ and $P \neq P_2$ in~$\tree$, i.e., the test of
line~\ref{new_line:separate_test} has failed.  Then, the test of
line~\ref{new_line:test_length} fails if and only if~$P$ is on the path
between~$P_1$ and~$P_2$ in~$\tree$.
\end{lemma}
\begin{proof} There are two directions.

$(\Longleftarrow)$ From lines~\ref{new_line:delta_one_two},
\ref{new_line:between_1} and 
Corollary \ref{cor_dist}(\ref{cor_dist_add}),
\begin{eqnarray*}
(\distsym'_1-\distsym'_2)-(\dist{P_1P}-\dist{P_2P}) & = &
\pm\left((\hatdist{X_1P}-\dist{X_1P})-
(\hatdist{X_1P_1}-\dist{X_1P_1})\right)
\\ 
& &
\pm\left((\hatdist{X_2P}-\dist{X_2P})-
(\hatdist{X_2P_2}-\dist{X_2P_2})\right).
\end{eqnarray*}
Thus, whether $P_1$ and $P_2$ are leaves or internal nodes in
$\hypo_k$, by $\AAA_k$, $\BB_k$, and $\centerevent$,
\(\left|(\distsym'_1-\distsym'_2)-(\dist{P_1P}-\dist{P_2P})\right| <
2\dist{\min}.
\) 
By line \ref{new_line:set_length_one} and Corollary
\ref{cor_dist}(\ref{cor_dist_add}),
\begin{eqnarray*}
\distsym''_1
& < & 
\frac{2\dist{\min}+(\dist{P_1P}-\dist{P_2P})+\dist{P_1P_2}^*}{2}
\\
&=& 
\frac{2(2\dist{\min}-\dist{P_2P})+(-2\dist{\min}+\dist{P_1P_2})+
\dist{P_1P_2}^*}{2}.
\end{eqnarray*}
Then, since $P \neq P_2$ and thus $\dist{P_2P} \geq 2\dist{\min}$, by
$\CC_k$, we have
\(
\distsym''_1 < \dist{P_1P_2}^*.
\)
By symmetry, $\distsym''_2 < \dist{P_1P_2}^*$.  Thus, the test of
line~\ref{new_line:test_length} fails.

$(\Longrightarrow)$ To prove by contradiction, assume that $P$ is not
on the path between~$P_1$ and~$P_2$.  By similar arguments, if
$\dist{P_1P}>\dist{P_1P_2}$ (respectively,
$\dist{P_2P}>\dist{P_1P_2}$), then $\dist{1}''>\dist{P_1P_2}^*$
(respectively, $\dist{2}''>\dist{P_1P_2}^*$). Thus, the test of
line~\ref{new_line:test_length} passes.
\end{proof}

\begin{figure}[h]
\centerline{
\psfig{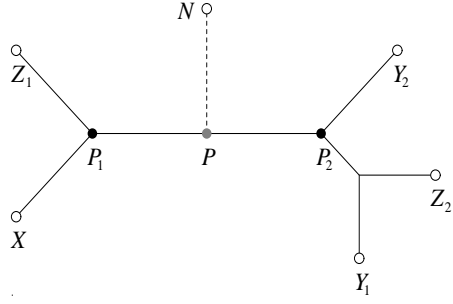}}
\caption{This subgraph of $\tree$ fixes some notation used in the
proof of Case 1 of Lemma \ref{new_lm:int_D}.  The location of $Y_1$
relative to $Y_2$ and $Z_2$ is nonessential; for instance, $Y_1$ can
even be the same as $Y_2$.  In $\hypo_k$, $\deftrip(P_1)=
\{X,Y_1,Z_1\}$ and $\deftrip(P_2)=\{X,Y_2,Z_2\}$. Neither $XY_1Z_1$
nor $XY_2Z_2$ is small, and $\dist{P_2Y_2}\le\dist{P_2Z_2}$.  We aim
to prove that there is a leaf~$N\not\in\hypo_k$ such that $NXY_2$ or
$NZ_1Y_2$ is large and defines a node $P$ strictly between $P_1$ and
$P_2$ in $\tree$.}
\label{new_fig:large_proof}
\end{figure}

\begin{lemma}\label{new_lm:int_D} 
Assume that $P_1P_2$ is an edge in $\hypo_k$ and some node is strictly
between $P_1$ and $P_2$ in $\tree$. Then there is a large
triplet~$NQ_1Q_2$ with center~$P$ such that~$N\not\in\hypo_k$,
$Q_1\in\deftrip(P_1)$, $Q_2\in\deftrip(P_2)$, and~$P$ is strictly
between~$P_1$ and~$P_2$ in $\tree$.
\end{lemma}
\begin{proof} 
By Lemma~\ref{fact:g-things}(\ref{fact:g:large}), for every node~$P$
strictly between~$P_1$ and~$P_2$ in~$\tree$, there exists a
leaf~$N\not\in \hypo_k$ with $\close{PN}\ge(1-\afact g)^{d+1}$.  To
choose $P$, there are two cases: (1) both $P_1$ and $P_2$ are internal
nodes in $\hypo_k$, and (2) $P_1$ or $P_2$ is a leaf in $\hypo_k$.

Case 1. By
Lemma~\ref{new_lem:cand_correct_endpoint},
let~$\deftrip(P_1)=\{X,Y_1,Z_1\}$ and~$\deftrip(P_2)=\{X,Y_2,Z_2\}$.
By $\BB_k$, neither $XY_2Z_2$ nor $XY_1Z_1$ is small.  To fix the
notation for $\deftrip(P_1)$ and $\deftrip(P_2)$ with respect to their
topological layout, we assume without loss of generality that
Figure~\ref{new_fig:large_proof} or equivalently the following statements
hold:
\begin{itemize}
\item 
In $\hypo_k$ and thus in $\tree$ by $\AAA_k$, $P_2$ is on the paths
between~$P_1$ and~$Y_2$, between~$P_1$ and~$Z_2$, and between $P_1$
and~$Y_1$, respectively.
\item 
Similarly, $P_1$ is on the paths between~$P_2$ and~$Z_1$ and
between~$P_2$ and~$X$.
\item 
$\dist{P_2Y_2}\le\dist{P_2Z_2}$.
\end{itemize}
Both~$NXY_2$ and~$NZ_1Y_2$ define $P$, and the target triplet is one
of these two for some suitable $P$. To choose $P$, we further divide
Case 1 into three subcases.

{\em Case} 1a: $\close{XP_2}<\close{Y_2P_2}(1-\afact g)$ and
$\close{Y_2P_1}<\close{XP_1}(1-\afact g)$.  The target triplet is
$NXY_2$.  Since $\close{XY_2}\leq\sqrt{\close{XY_2}}$, by
Corollary~\ref{cor_dist}(\ref{lem_somepoint}), 
let $P$ be a node on the path between
$X$ and $Y_2$ in $\tree$ with $\sqrt{\close{XY_2}(1-\afact
g)}\le\close{XP} \le\sqrt{\close{XY_2}(1-\afact g)^{-1}}$ and thus by
Lemma~\ref{lm:exp_close} $\sqrt{\close{XY_2}(1-\afact
g)}\le\close{Y_2P} \le\sqrt{\close{XY_2}(1-\afact g)^{-1}}$. By
the condition of Case 1a and Lemma~\ref{lm:exp_close},
$P$ is strictly between $P_1$ and $P_2$ in $\tree$. Also, by
Corollary~\ref{eq:avg}, $\close{XY_2}\ge\frac{2}{3}\close{XY_2Z_2}$.
Thus, by Lemma~\ref{lm:exp_close}, since $XY_2Z_2$ is
not small,
\begin{eqnarray}\label{new_eq:close_large_one}
\close{NXY_2}  & = &
\frac{3}{\frac{1}{\close{XP}\close{PN}}
        +\frac{1}{\close{Y_2P}\close{PN}}
        +\frac{1}{\close{XY_2}}}
\\ \nonumber
& \ge &
\frac{1}{\sqrt{\frac{2}{3}}\close{XY_2Z_2}^{-1/2}
(1-\afact g)^{-d-3/2}
+\frac{1}{2}\close{XY_2Z_2}^{-1}}
  >    \closelg.
\end{eqnarray}
So $NXY_2$ is as desired for Case 1a.

{\em Case} 1b: $\close{XP_2}\ge\close{Y_2P_2}(1-\afact g)$.  The
target triplet is $NXY_2$.  Let $P$ be the first node after $P_2$ on
the path from~$P_2$ toward $P_1$ in $\tree$.  Then,
$\close{Y_2P}\ge\close{Y_2P_2}(1-\afact g)$.  By
Corollary~\ref{eq:avg}, $\close{Y_2P}^2\ge\close{XY_2Z_2}(1-\afact
g)^2/3$.  Next, since $\close{XY_2}\ge\close{XZ_2}$ and
$\close{P_2Y_2}\ge\close{P_2Z_2}$,
\[
\close{XY_2Z_2}  
\leq  
\frac{3}{2\close{XY_2}^{-1}+\close{Y_2P_2}^{-1}\close{P_2Z_2}^{-1}} 
\leq  
\frac{3}{2\close{XP_2}^{-1}\close{Y_2P_2}^{-1}+
\close{Y_2P_2}^{-2}} 
\leq  
\frac{3\close{XP_2}^2}{2(1-\afact g)+(1-\afact g)^2}.
\]
So $\close{XP}^2 > \close{XP_2}^2 > \close{XY_2Z_2}(1-\afact g)^2$.
Since $\close{XY_2}\ge\frac{2}{3}\close{XY_2Z_2}$ and $XY_2Z_2$ is not
small,
\begin{eqnarray}\label{new_eq:close_large_two}
\close{NXY_2} & = & \frac{3}{\frac{1}{\close{XP}\close{PN}}
				+\frac{1}{\close{Y_2P}\close{PN}}
				+\frac{1}{\close{XY_2}}}
\\ \nonumber
& > &
\frac{1}{\left(\frac{1+\sqrt{3}}{3}\right)
\close{XY_2Z_2}^{-1/2}(1-\afact g)^{-d-2}
+ \frac{1}{2}\close{XY_2Z_2}^{-1}} 
  >   
\closelg.
\end{eqnarray}
So $NXY_2$ is as desired for Case 1b.  

{\em Case} 1c: $\close{Y_2P_1}\ge\close{XP_1}(1-\afact g)$.
If~$\close{Z_1P_1}>\close{XP_1}$, the target triplet is $NZ_1Y_2$;
otherwise, it is $NXY_2$.  The two cases are symmetric, and we assume
$\close{XP_1}\ge\close{Z_1P_1}$.  Let $P$ be the first node after
$P_1$ on the path from~$P_1$ toward~$P_2$ in $\tree$.  Then,
$\close{XP} \geq \close{XP_1}(1-\afact g)$. By Corollary~\ref{eq:avg},
$\close{XP}^2 \geq \close{XP_1}^2(1-\afact
g)^2\ge\close{XY_1Z_1}(1-\afact g)^2/3$.  Since
$\close{XY_2}\ge\close{XZ_2}$ and $\close{Y_2Z_2}>0$,
\[
\close{XY_2Z_2} \ < \ \frac{3}{2\close{XY_2}^{-1}} \ \le \
\frac{3}{2\close{Y_2P_1}^{-1}\close{XP_1}^{-1}} \ \le \
\frac{3\close{Y_2P_1}^2}{2(1-\afact g)}.
\]
Hence~$\close{Y_2P}^2 > \close{Y_2P_1}^2 >
2\close{XY_2Z_2}(1-\afact g)/3$.  Then, since neither $XY_2Z_2$ nor
$XY_1Z_1$ is small and 
$\close{XY_2}\ge\frac{2}{3}\close{XY_2Z_2}$,
\begin{eqnarray}
\label{new_eq:close_large_three}
& & \close{NXY_2}  =   \frac{3}{\frac{1}{\close{XP}\close{PN}}
				+\frac{1}{\close{Y_2P}\close{PN}}
				+\frac{1}{\close{XY_2}}}
\\ \nonumber
& > & \frac{1}{\frac{1}{\sqrt{3}}\close{XY_1Z_1}^{-1/2}
(1-\afact g)^{-d-2}
+\frac{1}{\sqrt{6}}\close{XY_2Z_2}^{-1/2}
(1-\afact g)^{-d-3/2}
+\frac{1}{2}\close{XY_2Z_2}^{-1}}
  >   \closelg.
\end{eqnarray}
So $NXY_2$ is as desired for Case 1c with
$\close{XP_1}\ge\close{Z_1P_1}$.

Case 2. By symmetry, assume that $P_2=X$ is a leaf in $\hypo_k$.
Since $k \geq 3$, $P_1$ is an internal node in $\hypo_k$.  Let
$\deftrip(P_1)=\{X,Y,Z\}$.  By symmetry, further assume
$\close{YP_1}\ge\close{ZP_1}$.  There are two subcases.
If~$\close{XP_1}<\close{YP_1}(1-\afact g)$, the proof is similar
to that of Case 1a and the desired $P$ is in the middle of the path
between~$X$ and~$Y$ in $\tree$.  Otherwise, the proof is similar that
of Case 1b and $P$ is the first node after $P_1$ on the path
from~$P_1$ toward~$X$ in $\tree$. In both cases, the desired triplet
is $NXY$.
\end{proof}

\subsection{The sample size required by \bfHGT}
\label{new_ss:main_sample}
The next lemma analyzes $\cand$. 
For $k=3,\ldots,\numtaxa-1$ and each leaf $M \in \tree$, let
$\cand_k[M]$ be the version of $\cand[M]$ at the start of the
$(k-2)$-th iteration of the repeat at line \ref{new_fhgt:repeat}.
\begin{lemma}\label{new_lm:cand_ok}
Assume that for a given $k \leq \numtaxa-1$, ~$\greedyevent$,
$\centerevent$, $\AAA_{k'}$,~$\BB_{k'}$, and~$\CC_{k'}$ hold for all
$k' \leq k$.
\begin{enumerate}
\item\label{new_f:null}
If $\cand_k[M]$ is not null, then it is a splitting tuple for
some edge in $\hypo_k$.
\item\label{new_f:leaf} 
If an edge $Q_1Q_2 \in \hypo_k$ and a triplet $MR_1R_2$ with
$M\not\in\hypo_k$ satisfy
Lemma~\ref{new_lm:int_D}, then $\cand_k[M]$
is a splitting tuple for $Q_1Q_2$ in $\hypo_k$ that contains
a triplet $MR'_1R'_2$ with
$\hatclose{MR'_1R'_2}\geq\hatclose{MR_1R_2}$.
\end{enumerate}
\end{lemma}
\begin{proof} The two statements are proved as follows.

Statement~\ref{new_f:null}. This statement follows directly from the
initialization of $\cand$ at line \ref{new_find_update_0}, the deletions
from $\cand$ at line \ref{new_find_update_2}, and the insertions into
$\cand$ at lines \ref{new_find_update_0} and \ref{new_find_update_3}.

Statement~\ref{new_f:leaf}. The proof is by induction on $k$.

{\it Base case}: $k = 3$.  By $\centerevent$, $\AAA_3$,~$\BB_3$,
$\CC_3$, and Lemmas \ref{new_lm:int_one} and
\ref{new_lm:int_two},
$MR_1R_2$ is a splitting triplet for $Q_1Q_2$ in
$\hypo_3$. By the maximization in Update-$\cand$ at
line~\ref{new_find_update_0}, $\cand[M]$ is a splitting tuple for
some edge $Q'_1Q'_2\in\hypo_3$ that contains a triplet $MR'_1R'_2$
with $\hatclose{MR'_1R'_2}\geq\hatclose{MR_1R_2}$.  By
$\greedyevent$, $MR'_1R'_2$ is not small.  By
Lemmas~\ref{new_lm:int_one} and \ref{new_lm:int_two}, $Q'_1Q'_2$ is $Q_1Q_2$.

{\it Induction hypothesis}: Statement~\ref{new_f:leaf} holds for
$k<\numtaxa-1$.

{\it Induction step}.  We consider how $\cand_{k+1}$ is obtained from
$\cand_{k}$ during the $(k-2)$-th iteration of the repeat at line
\ref{new_fhgt:repeat}.   There are two cases.

{\it Case} 1: $Q_1Q_2$ also exists in $\hypo_k$.  By
$\AAA_{k}$, $Q_1Q_2$ and
$MR_1R_2$ also satisfy
Lemmas~\ref{new_lm:int_one} and
\ref{new_lm:int_two} for $\hypo_k$.  By
the induction hypothesis, $\cand_k[M]$ is a splitting tuple
for $Q_1Q_2$ in $\hypo_k$ that contains a triplet $MR'_1R'_2$ with
$\hatclose{MR'_1R'_2}\geq\hatclose{MR_1R_2}$.  Then, since $Q_1Q_2 \neq
P_1P_2$ and $M\not=N$ at line \ref{new_find_update_2}, $\cand_k[M]$ is not
reset to null.  Thus, it can be changed only through replacement at
line \ref{new_find_update_3} by a splitting tuple for some edge
$Q'_1Q'_2$ in $\hypo_{k+1}$ that contains a triplet $MR''_1R'_2$ with
$\hatclose{MR''_1R''_2}\geq\hatclose{MR'_1R'_2}$.  By $\greedyevent$,
$MR''_1R''_2$ is not small.  Thus, by $\centerevent$, $\AAA_{k+1}$,
$\BB_{k+1}$, $\CC_{k+1}$, and
Lemmas~\ref{new_lm:int_one} and
\ref{new_lm:int_two}, $Q'_1Q'_2$ is
$Q_1Q_2$.

{\it Case} 2: $Q_1Q_2\not\in\hypo_k$.  This case is similar to the
base case but uses the maximization in Update-$\cand$ at
line~\ref{new_find_update_3}.
\end{proof}

\begin{lemma}\label{new_lm:sample_induct}
$\greedyevent$ and~$\centerevent$ imply that $\AAA_k$, $\BB_k$, and
$\CC_k$ hold for all~$k=3,\ldots,\numtaxa$.
\end{lemma}
\begin{proof} The proof is by induction on $k$.

{\it Base case}: $k=3$.  By
Lemma~\ref{fact:g-things}(\ref{fact:g:large}), $\centerevent$, and the
greedy selection of line~\ref{new_fhgt:init_triplet}, line
\ref{new_line:init_tree} constructs $\hypo_3$ without edge lengths.  Then,
$\AAA_3$ holds trivially.  $\BB_3$ follows from $\centerevent$,
$\greedyevent$, and line~\ref{new_fhgt:init_triplet}.  $\CC_3$ follows
from $\BB_3$, $\centerevent$ and the use of
Equation~\eqref{eq:est_center} at line~\ref{new_fhgt:init_length}.

{\it Induction hypothesis}: $\AAA_k$, $\BB_k$, and $\CC_k$ hold for
some $k < \numtaxa$.

{\it Induction step}.  The induction step is concerned with the
$(k-2)$-th iteration of the repeat at line \ref{new_fhgt:repeat}.  Right
before this iteration, by the induction hypothesis, since
$k<\numtaxa$, some $N'Q_1Q_2$ satisfies Lemma~\ref{new_lm:int_D}.
Therefore, during this iteration, by $\centerevent$ and
Lemmas~\ref{new_lm:int_one}, \ref{new_lm:int_two}, and
\ref{new_lm:cand_ok},
$\cand$ at line~\ref{new_hgt_fail_2} has a splitting tuple for $\hypo_k$
that contains a triplet $NXY$ with $\hatclose{NXY} \geq
\hatclose{N'Q_1Q_2}$.  Furthermore, line \ref{new_fhgt:choose} finds such
a tuple.  By $\greedyevent$, $NXY$ is not small.  Lines
\ref{new_line:add_int} and \ref{new_line:add_leaf} create $\hypo_{k+1}$ using
this triplet.  Thus, $\BB_{k+1}$ follows from $\BB_k$.  By
Lemmas~\ref{new_lm:int_one} and \ref{new_lm:int_two}, $\AAA_{k+1}$ follows
from $\AAA_k$.  $\CC_{k+1}$ follows from $\CC_k$ since the triplets
involved at line~\ref{new_line:set_length_one} are not small.
\end{proof}

\begin{theorem}\label{new_tm:sample}
For any $0<\delta<1$, using sequence length
\[
\smplen = \bigO{\frac{\log\frac{1}{\delta}+\log n}{(1-\afact
g)^{4d+8}f^2\accfactor^2}},
\]
\fHGT\ outputs $\hypo$ with the properties below with probability at
least~$1-\delta$:
\begin{enumerate}
\item Disregarding the edge lengths, $\hypo\sametopo\wtopo$.
\item 
For each edge $Q_1Q_2$ in $\hypo$, $|\dist{Q_1Q_2}^*-\dist{Q_1Q_2}|<
%\accfactor(-\ln(1-\afact f))$.
2\dist{\min}$.
\end{enumerate}
\end{theorem}

\begin{proof}
By Lemma~\ref{lem_events},
$\Prob{\overline\greedyevent}\le\frac{\delta}{2}$ if
\[
\smplen \ \ge \ 
\smplen_{\rm g} \ \eqdef \ 
210\afact^2 \frac{3\ln n+\ln\frac{3}{\delta}}{\closelg^2}.
\]
Similarly, by Lemma~\ref{lem_events},
$\Prob{\overline\centerevent}\le\frac{\delta}{2}$ if
\[
\smplen \ \ge \ 
\smplen_{\rm c} \ \eqdef \ 
81 \frac{3\ln n+\ln\frac{7}{\delta}}{\closelg^2 f^2\accfactor^2}.
\]
We choose $\smplen=\lceil\max\{\smplen_{\rm g},\smplen_{\rm
 c}\}\rceil$. Consequently, $\Prob{\greedyevent\mbox{ and
 }\centerevent}\ge 1-\delta$.  By Lemma~\ref{new_lm:sample_induct}, with
 probability at least~$1-\delta$, \fHGT\ outputs
 $\hypo_n$, and $\AAA_n$ and~$\CC_n$ hold, which correspond to the
 two statements of the theorem.
\end{proof}

\section{Further research}\label{sec_further}

We have shown that theoretically, \fHGT\ has the optimal time and
space complexity as well as a polynomial sample complexity.  It would
be important to determine the practical performance of the algorithm
by testing it extensively on empirical and simulated trees and
sequences.  Furthermore, as conjectured by one of the referees and
some other researchers, there might be a trade-off between the time
complexity and the practical performance.  If this is indeed true
empirically, it would be significant to quantify the trade-off
analytically.

\section*{Acknowledgments}
We thank Dana Angluin, Kevin Atteson, Joe Chang, Junhyong Kim, Stan
Eisenstat, Tandy Warnow, and the anonymous referees for extremely
helpful discussions and comments.  

\appendix

\section{Proofs of technical lemmas}

\subsection{Proof of Lemma~\ref{lm:est_center}}\label{app:lm:est_center}
Let~$h_{XY}=\frac{\hatclose{XY}}{\close{XY}}$;
$h_{XZ}=\frac{\hatclose{XZ}}{\close{XZ}}$;
$h_{YZ}=\frac{\hatclose{YZ}}{\close{YZ}}$.  By Equations
\eqref{eq:est_distance} and \eqref{eq:est_center}, and 
by conditioning on the events~$\{h_{XZ}\le 1-r\}$ 
and~$\{h_{YZ}\ge 1+s\}$ for some $r,s>0$,
\begin{eqnarray*}
&    & \Prob{\hatdist{XP}-\dist{XP} \ge \frac{-\ln(1-\epsilon)}{2}} 
  =    \Prob{{h_{XY}h_{XZ}} \le  {h_{YZ}}(1-\epsilon)}
\\
& \le & \Prob{h_{XZ}\le 1-r} + \Prob{h_{YZ}\ge 1+s}
		+ \Prob{h_{XY}\le(1-\epsilon)\frac{1+s}{1-r}}.
\end{eqnarray*}
Setting $\frac{1-r}{1+s} > 1-\epsilon$, by
Equations~\eqref{eq:close_fraction_less}
and~\eqref{eq:close_fraction_more},
\begin{eqnarray*}
& \Prob{\hatdist{XP}-\dist{XP} \ge \frac{-\ln(1-\epsilon)}{2}} 
 \le  &
\\
&                 \exp\left(-\frac{2}{\alpha^2}\smplen\close{XZ}^2r^2\right)
		+ \exp\left(-\frac{2}{\alpha^2}\smplen\close{YZ}^2s^2\right)
		+ \exp\left(-\frac{2}{\alpha^2}\smplen\close{XY}^2\left(
				1-(1-\epsilon)\frac{1+s}{1-r}\right)^2\right).
&
\end{eqnarray*}
Equating these exponential terms yields equations for~$r$ and~$s$. The
solution for~$r$ is
\[
r  =  \frac{t-\sqrt{t^2-u}}{2\close{XZ}\close{YZ}};\ 
t  =  \close{XY}\close{YZ} +\close{XZ}\close{YZ} 
	+(1-\epsilon)\close{XY}\close{XZ};\ 
u  = 4\close{XY}\close{YZ}^2\close{XZ}\epsilon.
\]
Using Taylor's expansion, for~$u>0$,
\(
(t-\sqrt{t^2-u})^2 > \frac{u^2}{4t^2}.
\)
Thus, 
\[
r^2 > \frac{\epsilon^2}{
	\left(\frac{1}{\close{XZ}}
			+\frac{1-\epsilon}{\close{YZ}}
			+\frac{1}{\close{XY}}\right)^2\close{XZ}^2}
	> \frac{\epsilon^2\close{XYZ}^2}{9\close{XZ}^2}.
\]
So
\(
\Prob{\hatdist{XP}-\dist{XP} \ge \frac{-\ln(1-\epsilon)}{2}} 
	 \le  3\exp\left(-\frac{2}{\alpha^2}\smplen\close{XZ}^2r^2\right) 
	 <  3\exp\left(-\frac{2}{9\afact^2}\smplen\close{XYZ}^2\epsilon^2\right).
\)
\subsection{Proof of Lemma~\ref{lm:greedy}}\label{app:lm:greedy}
We use the following basic inequalities.
\begin{equation}\label{eq:greedy_mm}
\min\left\{\frac{\hatclose{XY}}{\close{XY}},
			\frac{\hatclose{XZ}}{\close{XZ}},
			\frac{\hatclose{YZ}}{\close{YZ}}\right\} 
\le \frac{\hatclose{XYZ}}{\close{XYZ}} 
\le \max\left\{\frac{\hatclose{XY}}{\close{XY}},
			\frac{\hatclose{XZ}}{\close{XZ}},
			\frac{\hatclose{YZ}}{\close{YZ}}\right\};
\end{equation}
\begin{equation}\label{eq:greedy_min}
\frac{\close{XYZ}}{3} \le \min\left\{\close{XY}, \close{XZ}, \close{YZ}\right\}.
\end{equation}

The proof of Equation~\eqref{eq:greedy_small} is symmetric to that of
Equation~\eqref{eq:greedy_large}.  So we only prove
the latter.  Pick $\lambda\ge 1$
with $\close{XYZ}=\closelg\lambda$.  Without loss of generality,
we assume 
\(
\min\left\{\frac{\hatclose{XY}}{\close{XY}},
			\frac{\hatclose{XZ}}{\close{XZ}},
			\frac{\hatclose{YZ}}{\close{YZ}}\right\} 
	= \frac{\hatclose{XY}}{\close{XY}}.
\)
By Equations \eqref{eq:close_fraction_less},
\eqref{eq:greedy_mm}, and~\eqref{eq:greedy_min},
\begin{eqnarray*}
& &
\Prob{\hatclose{XYZ} \le \midpoint}  = 
\Prob{\frac{\hatclose{XYZ}}{\close{XYZ}}
\le \frac{\midpoint}{\closelg\lambda}}
	 \le  \Prob{\frac{\hatclose{XY}}{\close{XY}} \le
	\frac{\midpoint}{\closelg\lambda}}\\*
	& \le & \exp\left(-\frac{2}{\afact^2}\smplen\left(
				1-\frac{\midpoint}{\closelg\lambda}\right)^2\close{XY}^2\right)
	 \le  \exp\left(-\frac{2\left(1-\frac{\midpoint}{\closelg}\right)^2}{9\afact^2}\smplen
				\closelg^2\right).
\end{eqnarray*}
Then, Equation~\eqref{eq:greedy_large} follows from the fact that by
the choice of~$\midpoint$,
\[
\frac{2\left(1-\frac{\midpoint}{\closelg}\right)^2}{9\afact^2} 
	= \frac{(\sqrt{2}-1)^2}{36\afact^2}.
\]

\subsection{Proof of Lemma~\ref{lm:center}}\label{app:lm:center}
Since Lemma~\ref{lm:est_center}
can help establish only one half of the desired inequality, we split
the probability on the left-hand side of
Equation~\eqref{eq:center_ok}.
\begin{eqnarray*}
& & \Prob{\left|\hatdist{XP}-\dist{XP}\right| \ge \frac{\dist{\min}}{2}} 
\\*
	& \le & \Prob{\hatdist{XP}-\dist{XP} \ge \frac{\dist{\min}}{6}} + 
                \Prob{\hatdist{YP}-\dist{YP} \ge \frac{\dist{\min}}{6}} +  
\\*
	&&  
	    \Probc{\hatdist{XP}-\dist{XP} \le -\frac{\dist{\min}}{2}}{
			\hatdist{YP}-\dist{YP} < \frac{\dist{\min}}{6}}.
\end{eqnarray*}
Then, since 
\(
\hatdist{XY}-\dist{XY} = (\hatdist{XP}-\dist{XP})+(\hatdist{YP}-\dist{YP}),
\)
we have
\begin{eqnarray*}
& & \Probc{\hatdist{XP}-\dist{XP} \le -\frac{\dist{\min}}{2}}{
			\hatdist{YP}-\dist{YP} < \frac{\dist{\min}}{6}}
\\
& \leq &
\Prob{\hatdist{XY}-\dist{XY} \le -\frac{\dist{\min}}{3}}.
\end{eqnarray*}
Consequently, 
\begin{eqnarray}\label{eq:three_terms}
\Prob{\left|\hatdist{XP}-\dist{XP}\right| \ge \frac{\dist{\min}}{2}} 
	& \le &  \Prob{\hatdist{XP}-\dist{XP} \ge \frac{\dist{\min}}{6}} 
	+ 
\\ \nonumber & & \Prob{\hatdist{YP}-\dist{YP} \ge \frac{\dist{\min}}{6}} +
\\ \nonumber & & \Prob{\hatdist{XY}-\dist{XY} \le -\frac{\dist{\min}}{3}}.
\end{eqnarray}
By Lemma~\ref{lm:est_center},
\[
\Prob{\hatdist{XP}-\dist{XP} \ge \frac{\dist{\min}}{6}}
	 \le  3\exp\left(-\frac{2}{9\afact^2}\smplen\close{XYZ}^2\left(
		1-e^{-\frac{\dist{\min}}{3}}\right)^2\right).
\] 
By Taylor's expansion,
\(
\left(1-e^{-\frac{\dist{\min}}{3}}\right)^2 
	\ge \left(1-(1-\afact f)^\frac{\accfactor}{3}\right)^2 > \frac{\accfactor^2}{9}\afact^2f^2,
\) 
and thus
\begin{equation}\label{eq:center_term_one}
\Prob{\hatdist{XP}-\dist{XP} \ge \frac{\dist{\min}}{6}}	 
	\le 3\exp\left(-\frac{\accfactor^2}{81}\smplen\closelg^2 f^2\right).
\end{equation}
By symmetry,
\begin{equation}\label{eq:center_term_two}
\Prob{\hatdist{YP}-\dist{YP} \ge \frac{\dist{\min}}{6}}
	\le 3\exp\left(-\frac{\accfactor^2}{81}\smplen\closelg^2 f^2\right).
\end{equation}
By Equation~\eqref{eq:close_fraction_more},
\(
\Prob{\hatdist{XY}-\dist{XY} \le -\frac{\dist{\min}}{3}}
	\le \exp\left(-\frac{2}{\afact^2}\smplen\close{XY}^2\left(e^{\frac{\dist{\min}}{3}}-1\right)^2\right).
\)
{From} Equation~\eqref{eq:greedy_min},
\(
\close{XY}\ge\frac{\closesmall}{3\sqrt{2}}.
\)
By Taylor's expansion,
\(
\left(e^{\frac{\dist{\min}}{3}}-1\right)^2 \ge \left((1-\afact
	f)^{-\frac{\accfactor}{3}}-1\right)^2 > \frac{\accfactor^2}{9}\afact^2f^2.
\)
Therefore,
\begin{equation}\label{eq:center_term_three}
\Prob{\hatdist{XY}-\dist{XY} \le -\frac{\dist{\min}}{3}}
	\le \exp\left(-\frac{\accfactor^2}{81}\smplen\closelg^2 f^2\right).
\end{equation}
Lemma~\ref{lm:center} follows from the fact that putting
Equations~\eqref{eq:three_terms} through \eqref{eq:center_term_three}
together, we have
\(
\Prob{\left|\hatdist{XY}-\dist{XY}\right| \ge \frac{\dist{\min}}{2}}
	\le 7\exp\left(-\frac{\accfactor^2}{81}\smplen\closelg^2 f^2\right).
\)

\bibliographystyle{abbrv}
%\bibliography{evol,stat,all}
\bibliography{all}

\begin{thebibliography}{10}

\bibitem{Ag+96}
R.~Agarwala, V.~Bafna, M.~Farach, B.~Narayanan, M.~Paterson, and M.~Thorup.
\newblock On the approximability of numerical taxonomy $($fitting distances by
  tree metrics$)$.
\newblock {\em {SIAM} Journal on Computing}, 2000.
\newblock To appear.

\bibitem{adfk97}
A.~Ambainis, R.~Desper, M.~Farach, and S.~Kannan.
\newblock Nearly tight bounds on the learnability of evolution.
\newblock In {\em Proceedings of the 38th Annual IEEE Symposium on Foundations
  of Computer Science}, pages 524--533, 1997.

\bibitem{Att98}
K.~Atteson.
\newblock The performance of neighbor-joining algorithms of phylogeny
  reconstruction.
\newblock {\em Algorithmica}, 25(2-3):251--278, 1999.

\bibitem{CrGoGo98}
M.~Cryan, L.~A. Goldberg, and P.~W. Goldberg.
\newblock Evolutionary trees can be learned in polynomial time in the two-state
  general {Markov}-model.
\newblock In {\em Proceedings of the 39th Annual IEEE Symposium on Foundations
  of Computer Science}, pages 436--445, 1998.

\bibitem{kao.tree.dna.soda}
M.~Cs\H{u}r{\"{o}}s and M.~Y. Kao.
\newblock Recovering evolutionary trees through harmonic greedy triplets.
\newblock In {\em Proceedings of the 10th Annual ACM-SIAM Symposium on Discrete
  Algorithms}, pages 261--270, 1999.

\bibitem{csuros2000}
M.~Csuros.
\newblock {\em Reconstructing Phylogenies in {Markov} Models of Evolution}.
\newblock PhD thesis, Yale University, 2000.
\newblock Co-Directors: Dana Angluin and Ming-Yang Kao.

\bibitem{Day87}
W.~H.~E. Day.
\newblock Computational complexity of inferring phylogenies from dissimilarity
  matrices.
\newblock {\em Bulletin of Mathematical Biology}, 49:461--467, 1987.

\bibitem{DaJoSa86}
W.~H.~E. Day, D.~S. Johnson, and D.~Sankoff.
\newblock The computational complexity of inferring rooted phylogenies by
  parsimony.
\newblock {\em Mathematical Biosciences}, 81:33--42, 1986.

\bibitem{DuZhFe91}
D.-Z. Du, Y.-J. Zhang, and Q.~Feng.
\newblock On better heuristic for {Euclidean} {Steiner} minimum trees
  $($extended abstract$)$.
\newblock In {\em Proceedings of the 32nd Annual IEEE Symposium on Foundations
  of Computer Science}, pages 431--439, 1991.

\bibitem{essw-rsa}
P.~L. Erd{\H{o}}s, M.~A. Steel, L.~A. Sz{\'e}kely, and T.~J. Warnow.
\newblock A few logs suffice to build (almost) all trees. {I}.
\newblock {\em Random Structures \& Algorithms}, 14(2):153--184, 1999.

\bibitem{essw-tcs}
P.~L. Erd{\H{o}}s, M.~A. Steel, L.~A. Sz{\'e}kely, and T.~J. Warnow.
\newblock A few logs suffice to build (almost) all trees. {II}.
\newblock {\em Theoretical Computer Science}, 221(1-2):77--118, 1999.

\bibitem{faka96}
M.~Farach and S.~Kannan.
\newblock Efficient algorithms for inverting evolution.
\newblock {\em Journal of the {ACM}}, 46(4):437--449, 1999.

\bibitem{felsenstein82}
J.~Felsenstein.
\newblock Numerical methods for inferring evolutionary trees.
\newblock {\em The Quarterly Review of Biology}, 57:379--404, 1982.

\bibitem{felsenstein83}
J.~Felsenstein.
\newblock Inferring evolutionary trees from {DNA} sequences.
\newblock In B.~Weir, editor, {\em Statistical Analysis of DNA Sequence Data},
  pages 133--150. Dekker, 1983.

\bibitem{fel83}
J.~Felsenstein.
\newblock Statistical inference of phylogenies.
\newblock {\em Journal of the Royal Statistical Society Series A},
  146:246--272, 1983.

\bibitem{gus97}
D.~Gusfield.
\newblock {\em Algorithms on Strings, Trees, and Sequences: Computer Science
  and Computational Biology}.
\newblock Cambridge University Press, New York, NY, 1997.

\bibitem{Hoe63}
W.~Hoeffding.
\newblock Probability inequalities for sums of bounded random variables.
\newblock {\em Journal of the American Statistical Association}, 58:13--30,
  1963.

\bibitem{HNW1999}
D.~Huson, S.~Nettles, and T.~Warnow.
\newblock Disk-covering, a fast converging method for phylogenetic tree
  reconstruction.
\newblock {\em Journal of Computational Biology}, 6(3):369--386, 1999.

\bibitem{KMRRSS1994}
M.~J. Kearns, Y.~Mansour, D.~Ron, R.~Rubinfeld, R.~E. Schapire, and L.~Sellie.
\newblock On the learnability of discrete distributions (extended abstract).
\newblock In {\em Proceedings of the 26th Annual {ACM} Symposium on Theory of
  Computing}, pages 273--282, 1994.

\bibitem{Siddall1998}
M.~E. Siddall.
\newblock Success of parsimony in the four-taxon case: long-branch repulsion by
  likelihood in the {Farris} zone.
\newblock {\em Cladistics}, 14:209--220, 1998.

\bibitem{steel94}
M.~Steel.
\newblock Recovering a tree from the leaf colourations it generates under a
  {Markov} model.
\newblock {\em Applied Mathematics Letters}, 7(2):19--23, 1994.

\bibitem{SOWH96}
D.~L. Swofford, G.~J. Olsen, P.~J. Waddell, and D.~M. Hillis.
\newblock Phylogenetic inference.
\newblock In D.~M. Hillis, C.~Moritz, and B.~K. Mable, editors, {\em Molecular
  Systematics}, chapter~11, pages 407--514. Sinauer Associates, Sunderland, Ma,
  2nd edition, 1996.

\end{thebibliography}

\end{document}